\documentclass[baaa]{baaa}

\usepackage[pdftex]{hyperref}
\usepackage{subfigure}
\usepackage{natbib}
\usepackage{helvet,soul}
\usepackage[font=small]{caption}

\contriblanguage{1}

\contribtype{1}

\thematicarea{3}

\received{\ldots}
\accepted{\ldots}

\title{The Fe {\sc i} lines and the chromospheric activity}

\titlerunning{The Fe {\sc i} lines}

\author{
M. C. Vieytes\inst{1}
}

\authorrunning{Vieytes}

\contact{mariela@iafe.uba.ar}

\institute{
Instituto de Astronom{\'\i}a y F{\'\i}sica del Espacio, CONICET--UBA, Argentina
}

\resumen{Las líneas de Fe neutro son las más abundantes en el espectro de estrellas tardías. Se utilizan para obtener diferentes parámetros estelares básicos como abundancias relativas, gravedad superficial, y temperatura efectiva. Sin
embargo, los centros de muchas de estas líneas se forman en la cromósfera estelar, viéndose afectados por los cambios producidos por la actividad magnética. Calculando modelos semiempíricos de la atmósfera estelar fuera del equilibrio termodinámico local (NLTE por sus siglas en inglés), es posible estudiar la formación de estas líneas en detalle. En este trabajo
presentamos un nuevo modelo actualizado del átomo de Fe {\sc i}, que nos permite calcular 1715 líneas de Fe {\sc i} entre 3000 y 7000 \AA~ en estrellas con diferente nivel de actividad cromosférica. Los resultados
preliminares muestran la presencia de rangos espectrales más sensibles al calentamiento cromosférico, sugiriendo
que las líneas con mayores variaciones deben ser tenidas en cuenta como fuentes de error, e incluso deben ser excluídas, al momento de ser utilizadas para los fines descriptos anteriormente.}

\abstract{Neutral Fe lines are the most abundant in the spectrum of late stars. They are used to obtain different basic stellar parameters such as relative abundances, surface gravity, and effective temperature. 
However, the centers of many of these lines are formed in the stellar chromosphere, being affected by the changes produced by magnetic activity. Calculating non-local thermodynamic equilibrium (NLTE) semi-empirical models of the stellar atmosphere, it is possible to study the formation of these lines in detail. 
We present a new updated model of the Fe {\sc i} atom, which allows us to calculate 1715 lines of Fe {\sc i} between 3000 and 7000 \AA~in stars with different levels of chromospheric activity. 
Preliminary results show the presence of spectral ranges more sensitive to chromospheric heating, suggesting that the lines with greater variations must be taken into account as a source of uncertainty, and could even be excluded, when be used for the purposes described above.}

\keywords{stars: solar-type --- stars: chromospheres    --- stars: activity}

\begin{document}

\maketitle
\section{Introduction}\label{S_intro}
The Fe {\sc i} lines in the stellar spectra of late-type stars are used to characterize them. Basic parameters such as effective temperature, metallicity, or surface gravity can be calculated from these lines. They are also used in the detection of extrasolar planets by the radial velocity method. For these reasons, an understanding of their behavior under physical processes acting on line formation, such as the chromospheric activity, is crucial.  Any associate change due to these processes on the line profile can produce miscalculations if it is not properly understood and described. 

Previous observational studies have shown that some spectral Fe {\sc i} lines present changes in their depth and width due to chromospheric stellar activity \citep{wise18,spi20}. In some cases, the changes in the line profile can track the magnetic cycle \citep{liv07,flor16}.

Semi-empirical NLTE models of the stellar atmospheres can be used as a tool to study the formation of line profiles in detail. But to obtain reliable synthetic spectra, it is necessary to build atomic and molecular models of the species that form the atmosphere, considering the different processes that can take place between them and the radiation field. 

The goal of the present work is to update the data and expand the level structure of the Fe {\sc i} atomic model. Further on, using the new atomic model, we are able to calculate the spectra of a solar-type star at two different atmospheric thermal structures that differ in the level of chromospheric activity, which was estimated by the flux in the Ca K line profile. These synthetic spectra allow us to study the influence of activity on the Fe {\sc i} line profiles.

\section{the new Fe {\sc i} atomic model}
The previous Fe {\sc i} atomic model is described by \cite{fon15} and references therein. The energy level structure contained the first 119 energy levels, and their corresponding 381 sublevels from the National Institute of Standards and Technology (NIST) database version 2004 \citep{nist04}. The maximum energy reached using this atomic structure was 54386.189 cm$^{-1}$. This atomic model permitted the calculation, in NLTE, of 1624 Fe I lines. 

\begin{figure}[t]
\centering
\resizebox{\hsize}{!}{\includegraphics{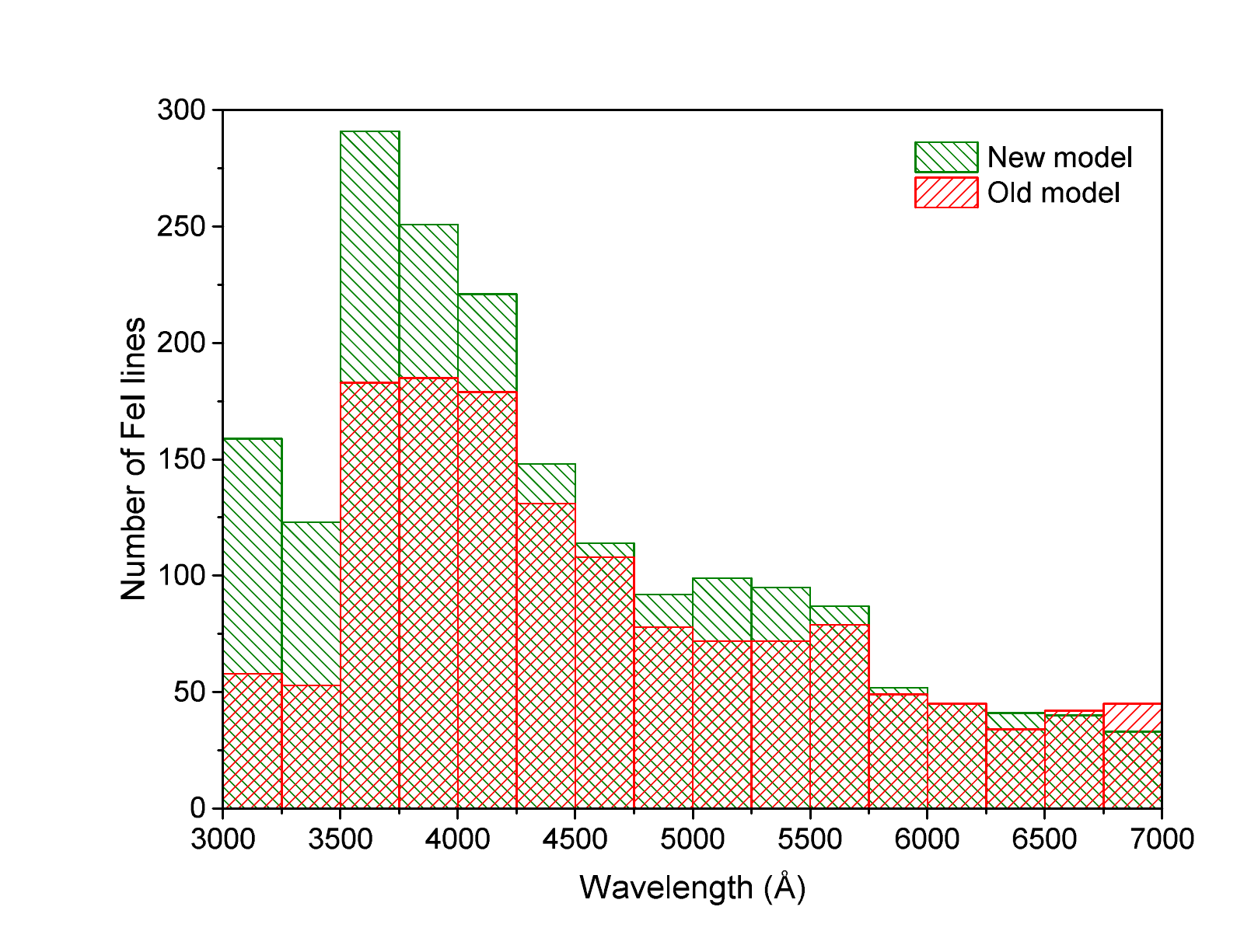}}
\caption{Comparison of the distribution of the Fe I lines calculated in NLTE with the new (green) and old (orange) atomic model between 3000 and 7000 \AA.}
\label{fig:histogram}
\end{figure}

We spanned and updated the atomic structure, using the latest version 2022 of the NIST database \citep{nist22}.The new Fe I model was built considering levels and sublevels until reaching a similar energy to the previous model, but the number of levels was increased to 141, and their corresponding sublevels to 444. 
We also selected from the same version of NIST, 2397 lines with well-determined oscillator strength, and 1715 lines in the range between 3000 and 7000 \AA. This portion of the spectra is mostly used to characterize the stars and for exoplanet detection. Considering that the previous atomic model included 1624 lines, we increased the calculated lines by 47\%. The distribution of these line centers in the visible spectrum for both atomic models is displayed in the histogram of Figure \ref{fig:histogram}. The new structure of energy levels and subleves allows to take into account an increased number of Fe {\sc i} lines mainly in the range between 3000 to 4250 \AA ~than the older model.

The line broadening processes for the line profiles considered were the usual, radiative, Stark, and van der Waals. The approximate values where taken from \cite{kurucz:1995} when the line was present, otherwise only the radiative coefficient was considered. 

Regarding the rest of the atomic parameters, the electron collisions were obtained from the semi-empirical approximation of \cite{sea62} for allowed transitions, and \cite{vanRege62}
when transitions were forbidden. For ionizing collisions,
the \cite{nrl05} ionization rate equation, adapted from Earth atmospheric calculations, was used. We neglected the H {\sc i} inelastic collisions because we didn't consider metal-poor stars.

The photoionization cross-sections were taken from TOPbase3,
the Opacity Project atomic database \citep{topbase93}. 
The radiative recombination was included as the
inverse of the photoionization process.  

\section{NLTE calculations}
There are several previous works that show the importance of an NLTE calculation for the level populations of Fe {\sc i}. The more comprehensive is the one by \cite{masho11}. \cite{hol13} showed that 3D NLTE calculation could be important for some regions with strong horizontal temperature gradients such as the boundaries of granules, but in spatial averaged profiles calculations, as is the case for stars, the difference between 1D and 3D NLTE is less than 1\%.

In this work, the calculations were done using the Solar-Stellar Radiation Physical Modeling ({\sc SSRPM}) system \citep{fon15}. This system allows us to self-consistently solve the equations of radiation transport and statistical equilibrium for a plane-parallel atmosphere, assuming hydrostatic equilibrium. 

The {\sc SSRPM} computes in NLTE the populations of H, H$^{-}$, H$_{2}$, and 52 neutral and lowly ionized atomic species, considering partial redistribution (PRD). These species are shown in Table 2 of \cite{fon15}, although in this work we modify the number of levels and sublevels of Fe {\sc i}. Additionally, the NLTE effectively thin approximation is used for highly ionized species. Besides, the most abundant diatomic molecular species are calculated in LTE. A more detailed explanation of the code can be found in \cite{fon16} and references therein. 

We selected the atmospheric model corresponding to the solar magnetic inter-network model 1401 built by \cite{fon15} to calculate the spectra representing a low chromospheric active G2 star. This feature is the most abundant in the atmosphere of the quiet Sun \citep{fon11}, and for that reason, it is a good model for a low-activity star. 

\begin{figure*}[ht]
\centering
\includegraphics[width=.5\textwidth]{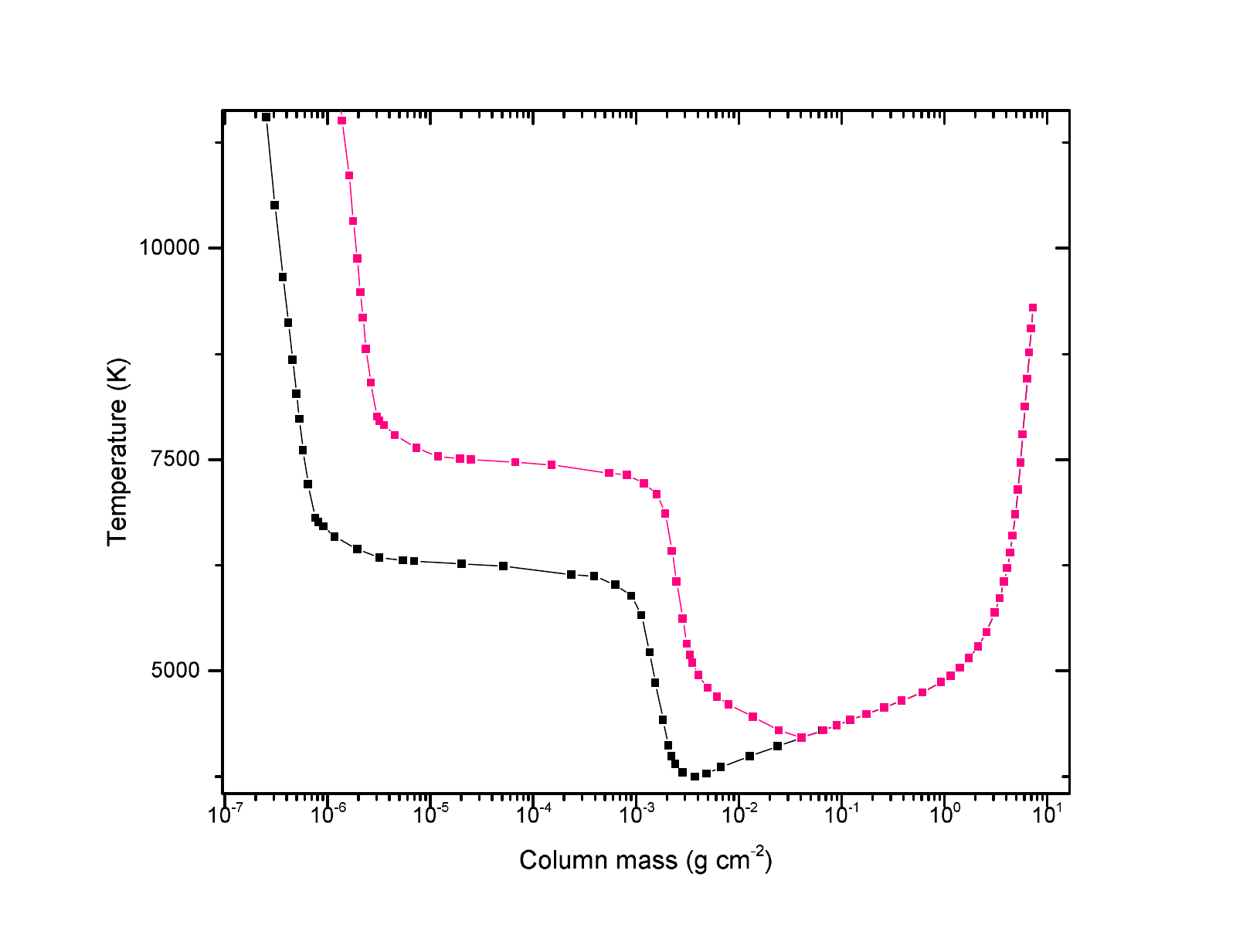}
\includegraphics[width=.5\textwidth]{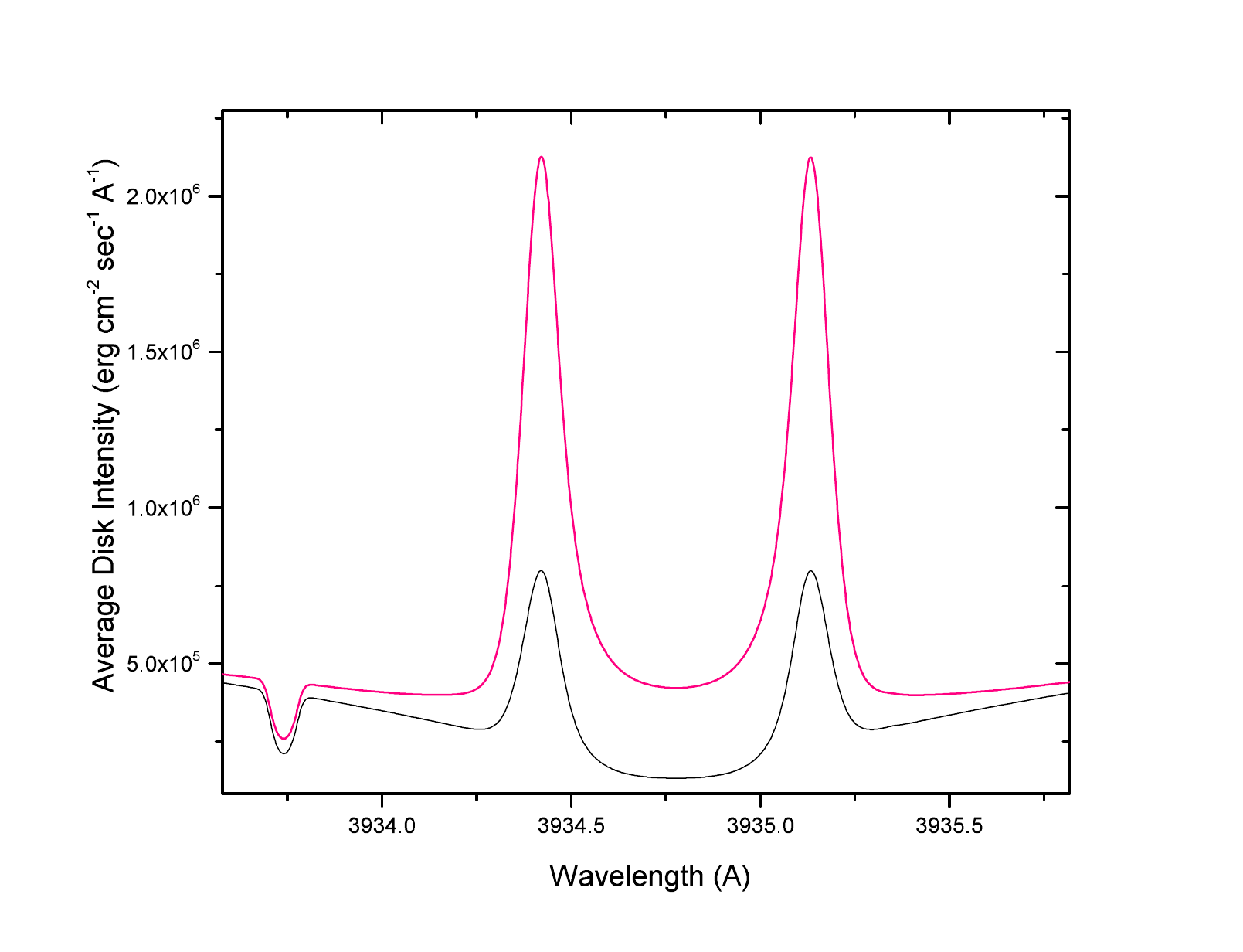}
\caption{\textbf{Left:} Atmospheric models with different chromospheric level of activity. In black the model for the Quiet Sun 1401, and in magenta a more active model built from it. \textbf{Right:} Ca II K profiles calculated from each model, plotted with the same line colour that the corresponding model. }
\label{fig:modelsyca2}
\end{figure*}

To generate a more chromospheric active G2 star, we used the same technique as in \citet{viey04} and \citet{andre95} by shifting the  thermal structure inward, i.e. toward higher mass column densities. Figure \ref{fig:modelsyca2} shows the thermal structure of the two models calculated, and their resulting Ca {\sc ii} K profiles. In black the model 1401, and in magenta a more active model built as it was previously described.

The Ca {\sc ii} H\&K doublet is widely used as a proxy of chromospheric activity due to its response to the chromospheric heating produced by stellar activity. For this reason, an increasing Ca {\sc ii} K profile ensures an increasing chromospheric activity.  It can be seen in the figure that the more active model gives a higher Ca {\sc ii} K profile, as is expected for a more chromospherically active star. 

It is important to note that both atmospheric models have the same photospheric thermal structure, as can be seen in Figure \ref{fig:modelsyca2}. We are interested only in studying the effect of a changing chromosphere due to the heating produced by the stellar activity. on the same star. 

\section{Results}
We computed the NLTE population for all the species compounding the atmosphere, and the resulting synthetic visible spectra. 

To determine the change in the Fe {\sc i} lines, we calculate the relative difference (in percentage) between both spectra. We identified the lines cross-crossing the wavelengths in the spectra with the line centers of the 1715 lines in our database belonging to the spectral range under study. 

\begin{figure}[t]
\centering
\resizebox{\hsize}{!}{\includegraphics{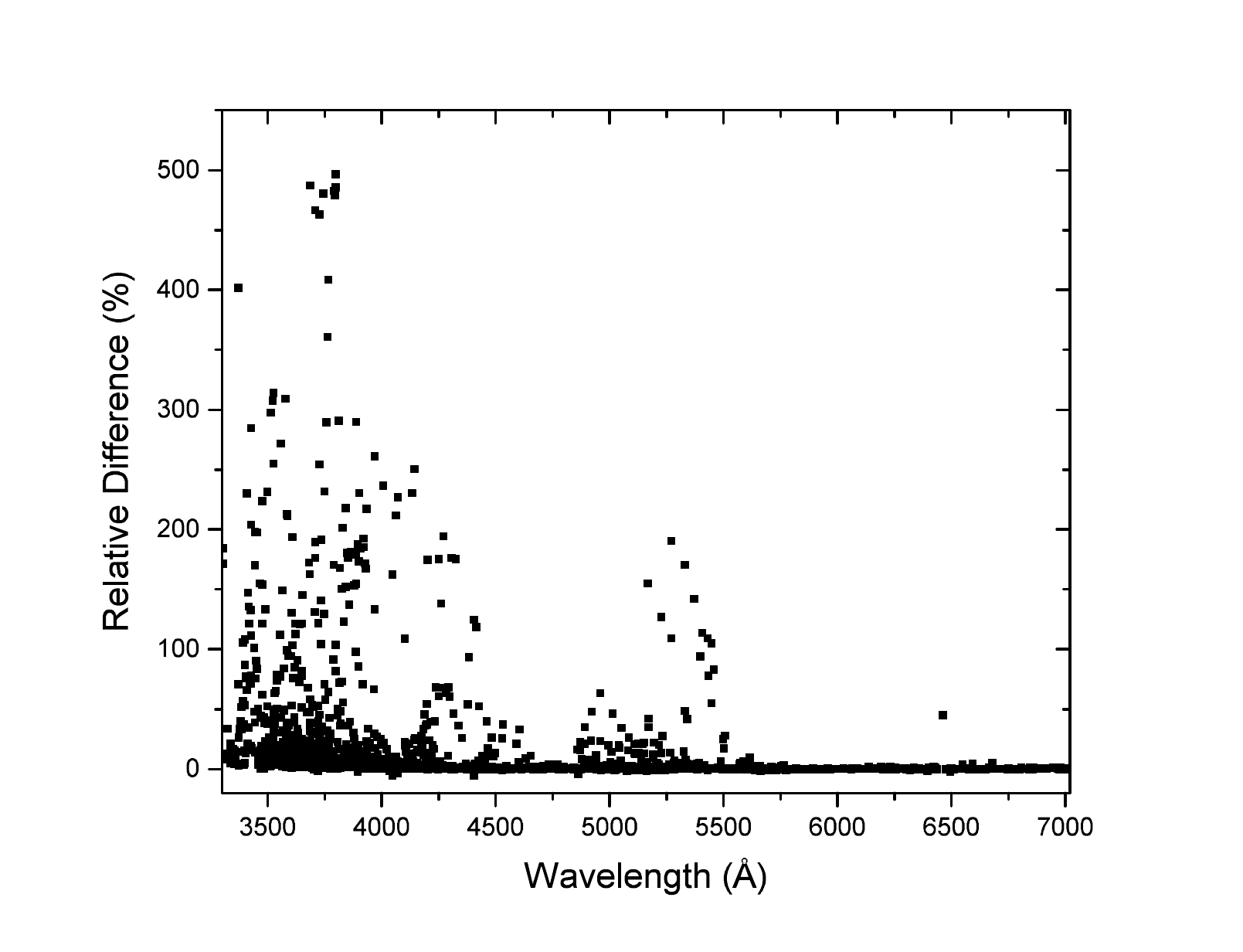}}
\caption{Relative difference in percentage between the Fe {\sc i} line centers calculated with the two models for the visible range.}
\label{fig:diferencias}
\end{figure}

Figure \ref{fig:diferencias} shows the result of the calculations for the spectral range from 3000 to 7000 \AA. 

38 \% of the Fe {\sc i} lines calculated with these models present a change in their line depths between 5 and 497 \%. 

It is possible to distinguish two main regions severally affected by the change in the thermal structure of the model, one between 3000 and 4550 \AA, and the other between 4780 to 5500 \AA ~approximately. 

The change shown in the figure means that these lines are affected in their line center by the level of chromospheric activity in the star. The main reason these lines are changing is that their formation region  move, totally or partially, outward to chromospheric regions as the temperature increases. 

62 \% of the lines are affected by less than 5 \%, showing that the peak of the contribution function remains at approximately the same photospheric height, where both models have the same thermal structure.

Although we were interested in studying the change in Fe {\sc i} lines with chromospheric activity, this methodology permits us to track the change in any line present in our synthetic spectra, for all the species from which the stellar atmosphere is made up. In a forthcoming paper, \cite{viey24} will address this idea, increasing the number of stars with different levels of chromospheric activity. 

\section{Conclusions}
Considering the importance of Fe {\sc i} in the stellar spectra of solar-type stars, we built a more comprehensive Fe {\sc i} atomic model. This new model increases 47\% of the number of lines that we can calculate with a 1D full NLTE semiempirical stellar model atmosphere. Using this new atomic model, we computed two atmospheric models representing dG2 stars with different levels of chromospheric activity. Calculating the relative difference in the percentage of the two synthetic spectra, we identified 652 (38 \%) Fe {\sc i} lines that are affected by chromospheric activity. These lines should be not considered in the stellar parameter calculations and be also excluded from the stellar mask used for exoplanet detections.

\begin{acknowledgement}
 This research was supported by grants PICT 2018-2895 and PICT 2019-4342 from the Agencia Nacional de Promoción Científica y Tecnológica (MINCyT, Argentina).
\end{acknowledgement}


\bibliographystyle{baaa}
\small
\bibliography{vieytes_AE}

\begin{thebibliography}{19}
\providecommand{\natexlab}[1]{#1}

\bibitem[{{Andretta} \& {Giampapa}(1995)}]{andre95}
{Andretta} V., {Giampapa} M.S., 1995, \apj, 439, 405

\bibitem[{{Cunto} et~al.(1993)}]{topbase93}
{Cunto} W., et~al., 1993, \aap, 275, L5

\bibitem[{{Flores} et~al.(2016)}]{flor16}
{Flores} M., et~al., 2016, \aap, 589, A135

\bibitem[{{Fontenla} et~al.(2015){Fontenla}, {Stancil} \& {Landi}}]{fon15}
{Fontenla} J.M., {Stancil} P.C., {Landi} E., 2015, \apj, 809, 157

\bibitem[{{Fontenla} et~al.(2011)}]{fon11}
{Fontenla} J.M., et~al., 2011, Journal of Geophysical Research (Atmospheres), 116, D20108

\bibitem[{{Fontenla} et~al.(2016)}]{fon16}
{Fontenla} J.M., et~al., 2016, \apj, 830, 154

\bibitem[{{Holzreuter} \& {Solanki}(2013)}]{hol13}
{Holzreuter} R., {Solanki} S.K., 2013, \aap, 558, A20

\bibitem[{Kramida et~al.(2004)}]{nist04}
Kramida A., et~al., 2004, {NIST Atomic Spectra Database (ver. 3.0.beta), [Online]. Available: {\tt{https://physics.nist.gov/asd}} [2004]. National Institute of Standards and Technology, Gaithersburg, MD.}

\bibitem[{Kramida et~al.(2022)}]{nist22}
Kramida A., et~al., 2022, {NIST Atomic Spectra Database (ver. 5.10), [Online]. Available: {\tt{https://physics.nist.gov/asd}} [2023, November 13]. National Institute of Standards and Technology, Gaithersburg, MD.}

\bibitem[{{Kurucz} \& {Bell}(1995)}]{kurucz:1995}
{Kurucz} R.L., {Bell} B., 1995, Atomic line data, kurucz cd-rom no. 23., \url{www.cfa.harvard.edu/amp/ampdata/kurucz23/sekur.html}. [Online; accessed: 2020, October 27]

\bibitem[{{Livingston} et~al.(2007)}]{liv07}
{Livingston} W., et~al., 2007, \apj, 657, 1137

\bibitem[{{Mashonkina} et~al.(2011)}]{masho11}
{Mashonkina} L., et~al., 2011, \aap, 528, A87

\bibitem[{{NRL Plasma Formulary}(2005)}]{nrl05}
{NRL Plasma Formulary} D.L., 2005, {NRL (Naval Research Laboratory) plasma formulary, revised}, Naval Research Lab. Report

\bibitem[{{Seaton}(1962)}]{sea62}
{Seaton} M.J., 1962, Proceedings of the Physical Society, 79, 1105

\bibitem[{{Spina} et~al.(2020)}]{spi20}
{Spina} L., et~al., 2020, \apj, 895, 52

\bibitem[{{van Regemorter}(1962)}]{vanRege62}
{van Regemorter} H., 1962, \apj, 136, 906

\bibitem[{{Vieytes} \& {Mauas}(2004)}]{viey04}
{Vieytes} M., {Mauas} P.J.D., 2004, \apss, 290, 311

\bibitem[{{Vieytes} et~al.(2024){Vieytes}, {Zhao} \& {Bedell}}]{viey24}
{Vieytes} M., {Zhao} L., {Bedell} M., 2024, in preparation

\bibitem[{{Wise} et~al.(2018)}]{wise18}
{Wise} A.W., et~al., 2018, \aj, 156, 180

\end{thebibliography}
 
\end{document}